\documentclass[nature,
,showpacs
,superscriptaddress
]
{revtex4-2}

\usepackage{newfloat,algcompatible}
\usepackage{etoolbox}
\AtBeginEnvironment{algorithm}{\noindent\hrulefill\par\nobreak\vskip-5pt}
\DeclareFloatingEnvironment[
    fileext=loa,
    listname=List of Algorithms,
    name=ALGORITHM,
    placement=tbhp,
]{algorithm}

\usepackage{graphicx,subfigure,bbm}

\usepackage{amsmath}
\usepackage{tabularx}
\usepackage{amsthm}
\usepackage{amssymb}
\usepackage{array,bm}
\usepackage{physics}
\usepackage{graphicx}
\usepackage{fancyhdr}
\usepackage{color}
\usepackage{extarrows}
\usepackage{enumerate}
\usepackage{epstopdf}
\usepackage[colorlinks,
            linkcolor=black,
            citecolor=black,
            urlcolor=blue
            ]{hyperref}
\usepackage[latin1]{inputenc}
\usepackage{tikz}
\usetikzlibrary{shapes,arrows}

\tikzstyle{decision} = [diamond, draw, fill=blue!20,
    text width=4.5em, text badly centered, node distance=3cm, inner sep=0pt]
\tikzstyle{block} = [rectangle, draw, fill=blue!20,
    text width=10em, text centered, rounded corners, minimum height=4em]
\tikzstyle{line} = [draw, -latex']
\tikzstyle{cloud} = [rectangle, draw,fill=red!20, node distance=7cm,
    minimum height=4em]

\newcommand{\be}{\begin{equation}}
\newcommand{\ee}{\end{equation}}
\newcommand{\bea}{\begin{eqnarray}}
\newcommand{\eea}{\end{eqnarray}}

\DeclareMathAlphabet{\varmathbb}{U}{bbold}{m}{n}

\newcommand{\argmin}{\mathop{\arg\min}}
\newcommand{\argmax}{\mathop{\arg\max}}
\newcommand{\kl}{D_{\mathrm KL}}

\begin{document}
\title{Generative Decoding for Quantum Error-correcting Codes}

\author{Hanyan Cao}
\affiliation{
CAS Key Laboratory for Theoretical Physics, Institute of Theoretical Physics, Chinese Academy of Sciences, Beijing 100190, China
}
\affiliation{
 School of Physical Sciences, University of Chinese Academy of Sciences, Beijing 100049, China
}

\author{Feng Pan}
\affiliation{
CAS Key Laboratory for Theoretical Physics, Institute of Theoretical Physics, Chinese Academy of Sciences, Beijing 100190, China
}
\affiliation{Centre for Quantum Technologies, National University of Singapore, 117543, Singapore}
\affiliation{Science, Mathematics and Technology Cluster, Singapore University
of Technology and Design, 8 Somapah Road, 487372 Singapore}

\author{Dongyang Feng}
\affiliation{
CAS Key Laboratory for Theoretical Physics, Institute of Theoretical Physics, Chinese Academy of Sciences, Beijing 100190, China
}
\affiliation{
 School of Physical Sciences, University of Chinese Academy of Sciences, Beijing 100049, China
}
\affiliation{School of Fundamental Physics and Mathematical Sciences, Hangzhou Institute for Advanced Study, UCAS, Hangzhou 310024, China}

\author{Yijia Wang}
\affiliation{
CAS Key Laboratory for Theoretical Physics, Institute of Theoretical Physics, Chinese Academy of Sciences, Beijing 100190, China
}
\affiliation{
 School of Physical Sciences, University of Chinese Academy of Sciences, Beijing 100049, China
}

\author{Pan Zhang}
\email{panzhang@itp.ac.cn}
\affiliation{
 CAS Key Laboratory for Theoretical Physics, Institute of Theoretical Physics, Chinese Academy of Sciences, Beijing 100190, China
}
\affiliation{School of Fundamental Physics and Mathematical Sciences, Hangzhou Institute for Advanced Study, UCAS, Hangzhou 310024, China}

\begin{abstract}
Efficient and accurate decoding of quantum error-correcting codes is essential for fault-tolerant quantum computation, however, it is challenging due to the degeneracy of errors, the complex code topology, and the large space for logical operators in high-rate codes. In this work, we propose a decoding algorithm utilizing generative modeling in machine learning. We employ autoregressive neural networks to learn the joint probability of logical operators and syndromes in an unsupervised manner, eliminating the need for labeled training data.
The learned model can approximately perform maximum likelihood decoding by directly generating the most likely logical operators for $k$ logical qubits with $\mathcal O(2k)$ computational complexity. Thus, it is particularly efficient for decoding high-rate codes with many logical qubits.
The proposed approach is general and applies to a wide spectrum of quantum error-correcting codes including surface codes and quantum low-density parity-check codes (qLDPC), under noise models ranging from code capacity noise to circuit level noise.
We conducted extensive numerical experiments to demonstrate that our approach achieves significantly higher decoding accuracy compared to the minimum weight perfect matching and belief propagation with ordered statistics on the surface codes and high-rate quantum low-density parity-check codes. 
Our approach highlights generative artificial intelligence as a potential solution for the real-time decoding of realistic and high-rate quantum error correction codes.
\end{abstract}

\maketitle

\section{Introduction}

Quantum computers have the potential to solve practical problems that are intractable for classical computers~\cite{factor1997,FTneed_2HowFactor20482021,quantumchemistry2005, vandamEndEndQuantumSimulation2024}.
However, current quantum computing devices are noisy, which restricts their capability. 
Quantum error correction (QEC) as a crucial step in managing these noises and achieving fault-tolerant quantum computing has become a key research frontier in both theoretical studies~\cite{Panteleev_2022, 10.1145/3519935.3520017, Bravyi_2024} and hardware developments~\cite{googlerep, google2023suppressing, googlenew2024} of quantum computation. 
In QEC, logical states comprising $k$ logical qubits are encoded using $n$ physical qubits, and continuous errors can be digitized into a finite set of discrete errors by measuring the redundant ancilla qubits, resulting in an error syndrome~\cite{Gottesman1997, nielsen_chuang_2010}. 
A decoding algorithm then interprets the error information based on the syndrome and determines an appropriate operation to correct the logical error.

Decoding in the quantum error correction codes is considered more challenging than in its classical counterpart due to the presence of multiple error types and their degeneracy~\cite{7336474}. 
Consequently, the corresponding Tanner graphs for quantum codes are more complex. 
For instance, the Tanner graph of Calderbank-Shor-Steane codes~\cite{PhysRevA.52.R2493, Calderbank_1996,PhysRevLett.77.793, Bravyi1998} invariably contains loops of various sizes due to commutation relations~\cite{7336474}. 
This complexity significantly diminishes the performance of standard decoding algorithms, such as belief propagation, in quantum systems compared to their classical counterparts.
Widely applied decoding algorithms include the minimum weight perfect matching (MWPM) algorithm~\cite{Dennis2001, higgott2021pymatching, wu2023fusion} and Belief propagation with ordered statistics post-processing (BPOSD)~\cite{412683, 924874, Roffe2020, Panteleev2021degeneratequantum}. Both methods fall within the minimum weight decoding category, which identifies the most probable error rather than the most probable logical error for given syndromes.

Another category of decoding algorithms, maximum likelihood decoding (MLD), addresses the issues by identifying the most likely logical operator among all $4^k$ possibilities.
In principle, MLD can provide optimal decoding results~\cite{Bravyi2014,chubb2021General, piveteau2023tensor,google2023suppressing}, however, this computational task belongs to the \#P-hard complexity class and remains highly challenging~\cite{Chubb_2021}.
Recently, efficient approximate implementations of MLD usually employ tensor networks, such as the boundary matrix product state method~\cite{Bravyi2014, chubb2021General} for surface code with code-capacity noise~\cite{Bravyi1998, Bombin_2012}. However, due to its high computational cost and the limited accuracy of approximate tensor-network contractions, tensor-network-based MLD poses significant challenges for quantum codes with complex topologies. 
For instance, this applies to three-dimensional codes~\cite{PhysRevA.100.012312}, codes with long-range interactions such as quantum low-density parity-check (qLDPC) codes~\cite{Panteleev_2022, 10.1145/3519935.3520017}, and codes subject to circuit-level noise. Another limitation of existing maximum-likelihood decoders is the requirement to compute probabilities for all $4^k$ logical operators corresponding to $k$ logical qubits, a task that becomes computationally infeasible for large $k$.

In this work, we introduce a neural MLD framework that overcomes the restrictions of arbitrary topology and the $4^k$ computational complexity of canonical maximum likelihood decoding methods. It is inspired by unsupervised generative modeling developed recently in machine learning~\cite{chung2014empiricalevaluationgatedrecurrent,germain2015made, Vaswani2017,6795963}. Our approach employs autoregressive neural networks to model the joint distribution of logical operators given a syndrome, using the product of conditional probabilities. The parameters of the neural network are learned from samples of the error model or experimental data without needing labels. The decoding process involves sequentially generating logical operators one by one for each logical qubit using the learned conditional probabilities, akin to generating chat text word by word in large language models~\cite{chatGPT}. We term it {\it generative neural decoder} (GND for short). Various autoregressive neural networks can be incorporated into our generative framework, including masked autoregressive network for density estimation (MADE)~\cite{germain2015made} and transformers~\cite{Vaswani2017, fakoor2020trade}.

In the past few years, various neural network (NN) decoders have been introduced~\cite{Varsamopoulos17, Krastanov2017,Baireuther2017,Overwater2022,Davaasuren2020,Gicev2021,PhysRevLett.128.080505}. These approaches typically frame decoding problems as classification tasks, utilizing neural networks such as multilayer perceptrons, convolutional neural networks, recurrent networks, and Transformers as classifiers to address the classification task.
Recently, some neural network decoders have been utilized to decode the surface code under the circuit-level noise ~\cite{ lange2023datadriven, bausch_learning_2024} and demonstrate significant advantages over classical decoding algorithms. However, classifier-based NN decoders usually work for a small number of logical qubits, particularly with $k=1$. Because they face a challenge that the number of classification labels scales exponentially as $4^k$~\cite{PhysRevLett.128.080505}. Some NN decoders resolve the $4^k$ complexity issue by treating $4^k$ logical classifications as $2k$ binary-classifications~\cite{Varsamopoulos17, Baireuther2017, lange2023datadriven}. Effectively, this approach employs an approximation known as the mean-field approximation to approximate the joint probability distribution of the $4^k$ logical operators using a product distribution, which approximation introduces additional errors.
In comparison, the generative decoding scheme proposed in this work naturally works for $k>1$ logical qubits and does not rely on the mean-field approximations, because it employs the autoregressive modeling~\cite{bishop2023deep} to represent the joint distribution as a product of conditional distributions. In other words, rather than simply mapping syndromes to error functions, our approach utilizes generative neural networks to model the joint distribution of logical operators given a syndrome, and sequentially generate the most probable logical operators, thereby circumventing the need for $4^k$ labels or $4^k$ computational complexity.

In the following text, we will first introduce our generative decoding scheme, then present numerical experiments on the surface code~\cite{Bravyi1998, Bombin2007}, Bivariate Bicycle (BB) code~\cite{Bravyi_2024}, and qLDPC codes~\cite{qldpc} under various error models to demonstrate the superiority of our generative decoding approach.

\section{Results}
\subsection{The generative decoding scheme}
\begin{figure*}[!htbp]
\centering
\includegraphics[scale=1.1]{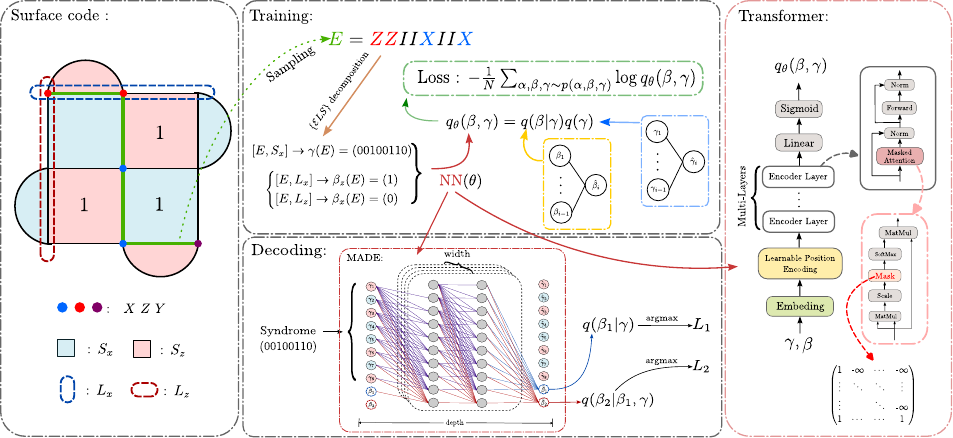}
\caption{Illustration of GND's training and decoding processes, using a distance-3 surface code as an example. The red and blue plaquettes represent the X and Z stabilizer generators, respectively, acting on the qubits located at the lattice vertices they cover. The training process begins by sampling errors from the error model, which are then digitalized using the $\mathcal{ELS}$ decomposition to obtain the corresponding configurations. These configurations are fed as input to the autoregressive neural network, and the output of the network is used to compute the negative log-likelihood (NLL), followed by a gradient descent optimization step. The observed syndrome is fed into the neural network during decoding to obtain the first logical variable. This variable is then fed back into the network along with the syndrome until the final logical variable is determined. The figure also illustrates the specific structures of the two autoregressive neural networks employed in this work, MADE and causal Transformer.} 
\label{fig:1}
\end{figure*}

Consider a $[[n,k,d]]$ quantum code, where $k$ logical qubits are encoded using $n$ physical qubits, and the code distance is $d$. A state $|\psi\rangle$ in the logical space is encoded as a fixed point of operators $\{S\}$ with eigenvalue equal to $1$, i.e. $S\ket{\psi}=\ket{\psi}$. 
These operators form an Abelian subgroup of the $n$-qubit Pauli group (ignoring the global phase) $\mathcal P_n=\{I,X,Y,Z\}^{\otimes n}$, known as the stabilizer group. Apparently, applying them to the quantum circuit does not influence the encoded states. This group $\mathcal{S}=\langle g_1,g_2,\cdots,g_m\rangle$ has $m=n-k$ generators, referred to as stabilizer generators in the stabilizer formalism~\cite{Gottesman1997, nielsen_chuang_2010}.
When an error occurs on the state $\ket{\psi}$, it is modeled as the application of an error operator $E \in \mathcal{P}_n$. Most errors do not commute with the stabilizers, yielding $-1$ eigenvalue upon measurement. As illustrated in Fig.~\ref{fig:1}, these errors can be detected by measuring the ancilla qubits associated with the stabilizer generators, producing a syndrome $\gamma(E)=\{\gamma_1(E),\gamma_2(E),\cdots, \gamma_m(E)\}$, where $\gamma_i(E)=0$ if $g_i$ and $E$ commute, and $\gamma_i(E)=1$ if they anti-commute.

Whether and how the error $E$ affects the encoded states depends on its associated logical operation. For instance, a quantum code encoding a single logical qubit has four possible logical operators $L_I, L_X, L_Z, L_Y$. An error $E$ belongs to one of the four corresponding logical sectors, determined by its commutation relations with the logical operators. 
By ignoring the global phase, we only need to consider $L_X$ and $L_Z$, the $4$ logical sectors are characterized by commutation relations with these two operators. This can be represented using $2$ binary variables $\beta=\{\beta^X(E)$ and $\beta^Z(E)\}$, indicating whether $E$ commute with $L_X$ and $L_Z$. 
The formulation above can be generalized to codes with $k$ logical qubits with $4^k$ logical operators, by using $2k$ binary variables for logical sectors $\{\beta_1,\cdots,\beta_k\}=\{\beta_1^X(E),\beta_1^Z(E),\cdots,\beta_k^X(E),\beta_k^Z(E)\}$. 
Then we can see that the task of decoding is to categorize logical sectors $\beta$ given syndrome $\gamma$, that is, model the conditional probability $p(\beta|\gamma)$ with $\beta\in \{0,1\}^{2k}$ and $\gamma\in\{0,1\}^m$. 
Given an error model, samples of the joint distribution $p(\beta,\gamma)$ can, in principle, be directly drawn from the model. However, accurately computing the conditional probability distribution 
$$p(\beta|\gamma)=\frac{p(\beta,\gamma)}{\sum_{\beta}p(\beta,\gamma)}$$ 
for a large number of logical qubits is challenging, as naively computing the normalization factor (i.e., the partition function in statistical physics) has a computational complexity of $2^{2k}$. 
In the exsiting neural network decoders\cite{8880492, Krastanov2017, PhysRevLett.119.030501, PhysRevLett.122.200501, bausch_learning_2024}, the number of logical qubits $k$ is usually considered to be small (particularly $k=1$ in the surface code) and thus this challenge is often ignored.

In this work, we propose to approach $p(\beta,\gamma)$ by learning a structured variational distribution known as \textit{autoregressive model}. The model factorizes the variational joint distribution $q_{\theta}(\beta,\gamma)$ into the product of conditional probabilities.
\begin{equation}
q_{\theta}(\beta,\gamma)=\prod_{i=1}^{2k} q(\beta_i|\beta_{j<i},\gamma)\cdot \prod_{i=1}^m q(\gamma_i|\gamma_{j<i})\;.
\label{eq:1}
\end{equation}
Here, the subscript $j<i$ denotes variables in front of variable $i$ given an order of variables. The $i$'th logical variable $\beta_i$ depends solely on the syndrome $\gamma$ and logical variables before it $\beta_{j<i}$. The order of variables is given naturally in this case and can be considered as a predefined arrow of time. in this sense, the variables in front of $i$ are the history of $i$, and variables behind of $i$ are $i$'s future. So the states of $i$ only depend on its history $j<i$, not on its future $j>i$. This restriction is essential in autoregressive modeling and is sometimes called ``casual''. Autoregressive models encompass various renowned implementations, including recurrent neural networks (RNNs)~\cite{chung2014empiricalevaluationgatedrecurrent}, long short-term memory (LSTM)~\cite{6795963}, and causal transformers~\cite{Vaswani2017, fakoor2020trade}, all of which can be integrated into our GND framework. 
In this work, we consider two concrete examples of autoregressive networks: the masked autoregressive network for density estimation (MADE)~\cite{germain2015made}, which has applications in statistical mechanics~\cite{PhysRevLett.122.080602}, and the Transformer~\cite{Vaswani2017, fakoor2020trade}, a standard model in natural language processing for generating subsequent words based on a given sentence. Their structures are illustrated in Fig.\ref{fig:1}.

The autoregressive model $q_\theta$ is trained to approximate the data distribution $p(\beta,\gamma)$. This is achieved by learning the parameters $\theta$ to minimize the distance between the
$p(\beta,\gamma)$ and $q_\theta(\beta,\gamma)$. We assume that the samples of the true distribution $p(\beta,\gamma)$ can be obtained efficiently, either using an error model or from experiments.
In this work, we choose the forward Kullback-Leibler divergence 
\begin{equation}
 \kl(p \| q_\theta)=
\sum_{\beta,\gamma} p(\beta,\gamma) \log \left[ \frac{p(\beta,\gamma)}{q_\theta(\beta,\gamma)}\right ],
\label{eq:2}
\end{equation}
as the distance measure, this yields a negative log-likelihood loss function
\begin{equation}
 \mathcal L=
 \argmin_\theta -\sum_{\alpha,\beta,\gamma} p(\alpha,\beta,\gamma)\log q_\theta(\beta,\gamma).
\label{eq:3}
\end{equation}
Here, $\alpha$ denotes the configuration of stabilizer generators, as detailed in Section \ref{els}. Directly optimizing $\mathcal L$ using Eq.\ref{eq:2} requires samples drawn from $p(\beta, \gamma)$, which is challenging since $p(\beta, \gamma) = \sum_{\alpha} p(\alpha, \beta, \gamma)$. Fortunately, as Eq.\ref{eq:3} indicates, we can directly sample from the error model $p(\alpha, \beta, \gamma)$ to train the autoregressive model, eliminating the need to prepare datasets of $(\beta, \gamma)$.

The training process is depicted in Fig.~\ref{fig:1}. First, errors are sampled from the error model, and the $\{\mathcal {E},\mathcal {L},\mathcal{S}\}$ decomposition is applied to each error to obtain the $(\beta, \gamma)$ configuration, which serves as training data for the autoregressive neural network. Two specific autoregressive neural networks, the masked autoregressive network for density estimation (MADE)~\cite{germain2015made} and the Transformer~\cite{Vaswani2017, fakoor2020trade}, are illustrated.
To facilitate learning, we iteratively minimize the loss function in Eq.~\eqref{eq:3} to reduce the discrepancy between the autoregressive variational distribution and the error distribution. After training, the autoregressive network learns all conditional probabilities as specified in Eq.~\eqref{eq:1}. Decoding proceeds by sequentially generating a logical operator configuration $\beta=\{0,1\}^{2k}$ using the learned conditional probabilities conditioned on the syndrome.

\begin{equation}\label{eq:4}
\hat \beta_i  = \argmax_{\beta_i} q_{\theta}(\beta_i|\beta_1,\cdots,\beta_{i-1},\gamma_1,\cdots,\gamma_{m})\;,
\end{equation}

This process is analogous to text generation by ChatGPT~\cite{Radford2018ImprovingLU,chatGPT,openai2023gpt4}, where text is generated word-by-word given a prompt (which is analogous to syndromes in our setting). In more detail, decoding the $i$th logical variable involves setting the syndrome $\gamma$ and the logical variable $\beta_{<i}$ as input (other variables are set to zero to maintain consistent input length) and running the forward pass of the neural network to obtain the output, i.e. the conditional probability $p(\beta_i|\gamma,\beta_{j<i})$, then decide the logical operator for $i$ using Eq.~\eqref{eq:4}. 
So for each logical operator, it takes a forward pass of the neural network, so in total requires $2k$ neural network passes for decoding $k$ logical qubits, contrasting sharply with the $\mathcal{O}(4^k)$ complexity of the conventional tensor network-based MLD method. Also, our scheme is capable of handling arbitrary code topologies, easily accommodating any stabilizer generator connectivity.

\subsection{Numerical experiments on high-rate quantum codes}

\begin{figure*}[!htb]
\centering
\includegraphics[width=\linewidth]{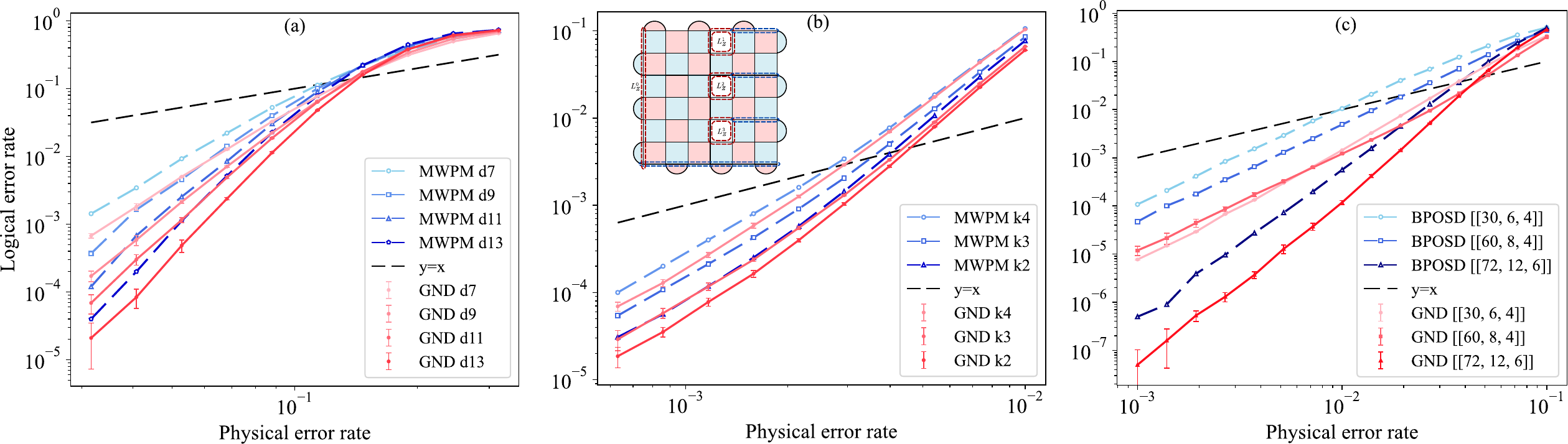}
\caption{Performance of our generative neural decoder (GND) compared with the minimum weight perfect matching (MWPM) and the BPOSD algorithm. (a) Rotated surface code with distances up to 13 under depolarizing noise. (b) Defected surface code originally of distance 7. We introduce 3 defects by removing 3 $Z$-stabilizer generators. Each defect provides an additional logical qubit, for which we establish a valid logical $Z$ operator base represented by red dashed lines. The minimum weight of a logical $Z$ operator is 4, corresponding to the Z-distance. The blue dashed lines represent the corresponding X-operators $L_X$. (c) High-rate [[60, 8, 4]], [[72, 12, 6]] BB code and [[30, 6, 4]] qLDPC codes under code capacity depolarizing noise model. We used the \textsc{Pymatching}~\cite{higgott2021pymatching} for MWPM decoding and the \textsc{ldpc} package~\cite{Roffe2020} for BPOSD decoding (using min-sum BP, the max number of iterations is 1000, and the post-processing method is OSD-CS). 
For GND, the neural networks are trained at a single physical error rate equal (or close) to the threshold and used for decoding at all physical error rates. For rotated surface codes and [[30, 6, 4]] ldpc code, we used $2\times10^9$ samples to train $2\times10^5$ epochs using 10000 batch; for [[60, 8, 4]] BB codes and defected surface codes under curcuit level noise, we used $5\times10^9$ data to train $5\times10^5$ epochs using 10000 batch; for [[72, 12, 6]] BB code, we used $1\times10^{10}$ data to train $5\times10^5$ epochs using 20000 batch. Each data point in (a) and (b) is averaged over 100
decoding tasks with each task comprising $10^5$ error instances, and in (c) is averaged over 10 decoding tasks, with each task involving $10^7$ error instances.
} 
\label{fig:3}
\end{figure*}
To evaluate our generative decoding approach, we conducted extensive numerical experiments on several quantum codes with $k=1$ and $k>1$ logical qubits. We tested both MADE and Transformer as autoregressive networks within our GND framework. Although Transformer has demonstrated superior performance in language modeling~\cite{chatGPT}, it did not exhibit greater capability than MADE in our framework. Consequently, in the main text, we report only the results obtained with MADE. A detailed comparison between MADE and Transformer is provided in the Sec~\ref{sec:Iv.c}.

We first evaluate GND on the standard rotated surface code (with $k=1$) under the depolarizing noises. The depolarising noise only considers the equal probability errors on the qubit and simplifies the decoding problem, so we can test GND with a distance up to $13$. The comparisons between GND and the MWPM algorithms are shown in Fig.~\ref{fig:3}. We can see that at any physical error rate, GND gives a lower logical error rate than MWPM, and the improvements are quite significant at low physical error rates.

Next, we examine the codes with $k>1$ logical qubits. We first choose the defected surface code under the circuit-level noise. As a variation of the surface code, the defected surface code removes several stabilizers to achieve a higher code rate~\cite{PhysRevA.86.032324}. This code retains the two-dimensional structure of the surface code and is well-suited for widely used superconducting quantum chips. Experiments on the defected surface code can be conveniently conducted with only minor modifications to the conventional surface code setup, thus serving as an effective method to evaluate the efficiency of high-rate codes in near-term applications. For the noise, we consider the circuit-level noise model, which is more experimentally realistic than the code capacity noise considered in the last example. The circuit-level noise accounts for errors in actual quantum circuits, including measurements, gate operations, idling, and reset~\cite{google2023suppressing,googlenew2024,2022yimingsurfacerealization,gaugephd}. To correct such noise, repetitive syndrome measurements are required~\cite{bohdanowicz2022quantum}, which often introduces redundant information. A more compact representation is provided by the detector error model (DEM)~\cite{McEwen2023relaxinghardware}, where a detector is defined as the linear exclusive-or (XOR) of multiple measurements. DEM captures discrepancies in consecutive measurements across circuit layers, offering a more concise representation than full syndrome measurement results. In this framework, detector states serve as syndromes $\gamma$, while final logical errors correspond to $\beta$ in our method. 

As illustrated in the inset of Fig.~\ref{fig:3}, we consider a distance-7 surface code with three stabilizers removed. The defected surface code features one Z-operator $L^0_{Z}$ of length 7 and three $Z$-operators $L^1_Z, L^2_Z, L^3_Z$ of length 4. Consequently, it encodes 4 logical qubits and has a $Z$-distance of 4.
To evaluate decoding performance across different $k$ values, we sequentially removed the stabilizers corresponding to $L^1_Z, L^2_Z, L^3_Z$, resulting in three defected surface codes with $k = 2, 3, 4$. 
In the numerical experiments, we initialized the circuit in the $Z$-basis and performed four rounds of syndrome measurements. We utilize the Stim~\cite{gidney2021Stim} library to construct the DEM, simulate the circuit, and perform sampling. The results, shown in Fig.~\ref{fig:3}(b), compare the logical error rates of our method GND and MWPM for each $k$ value. The figures demonstrate that GND achieves significantly lower logical error rates than MWPM for all $k$ values across all physical error rates.

Although the surface code has a two-dimensional structure that is convenient to the existing superconducting quantum hardware, it has the disadvantage of la ow code rate $R=k/n$, the ratio between the number of logical qubits and the number of physical qubits. So it is inefficient to encode a large number of logical qubits. Recently, researchers investigated quantum codes that have a much larger logical rate than the surface code; here we term them as ``high-rate codes''. Notable examples include Bivariate Bicycle (BB) code~\cite{Bravyi_2024}, and qLDPC codes~\cite{qldpc}. The BB codes family, as a specialized class of qLDPC codes, exhibits several advantageous properties. First, compared to commonly used $k=1$ surface codes, it significantly reduces physical qubit resources while encoding the same number of logical qubits with equivalent accuracy. Additionally, it features a well-designed syndrome measurement circuit with only a few layers of parallel two-qubit gates, maintaining circuit distance and preventing hook errors. Furthermore, due to its specially designed topology, it can be embedded into a two-layer quantum hardware architecture with efficient long-range connectivity, making it a promising candidate for future fault-tolerant quantum computing. To further demonstrate the universality of our GND method for arbitrary topology high-rate codes, we also evaluated small qLDPC codes. The codes are chosen after a random search for those lacking a regular topological structure, making them ideal candidates for our evaluation.
In our numerical experiments, we take the [[60, 8, 4]] and [[72, 12, 6]] BB codes, along with a [[30, 6, 4]] qLDPC code under code capacity depolarizing noise model as illustrative examples to demonstrate the performance of GND.

We note that in exchange for higher encoding rates, the BB codes and qLDPC codes exhibit more complex structures, with each qubit checked by multiple stabilizers. Consequently, when mapping the decoding problem to a matching problem, hyperedges emerge, rendering MWPM inapplicable. Thus, we compare our algorithm with the only currently available decoding algorithm, BPOSD~\cite{412683, 924874, Roffe2020, Panteleev2021degeneratequantum}. The comparison results are shown in Fig.~\ref{fig:3}(c). The results demonstrate that our method surpasses BPOSD in logical error rates for both BB codes and the selected qLDPC code. With lower error rates, the gap between GND and BPOSD is larger. Notably, with a physical error rate below $10^{-2}$, GND is about $10$ times more accurate than BPOSD.

\section{Discussions}

We have presented a neural network decoding framework for quantum codes based on generative modeling in machine learning. In this framework, an autoregressive generative model is employed to approximate the conditional probability of the target logical sector given the syndrome (detector), and trained by minimizing the discrepancy (measured by KL divergence) between the empirical error distribution and the distribution of the autoregressive model. After training, it can generate logical operators sequentially using the learned conditional probabilities, analogous to generating the text in the large language models. 
Our approach shares advantages common to other NN decoders, such as applicability to arbitrary code topologies, arbitrary error models, and fast neural network computation. The significant advantage of our approach is to avoid computing $4^k$ probabilities for logical operators.
This is because our model factorizes the variational joint distribution encompassing $4^k$ configurations into a product of $2k$ conditional probabilities. This greatly reduces the generation of logical operators for $k>1$ logical qubits, and the number of neural network computations required for decoding scales linearly with the number of logical qubits. 
Furthermore, compared to existing NN decoding schemes, our approach avoids using exponentially large classification labels or using product-of-marginal approximation. We have carried out extensive numerical experiments using surface codes, the defected surface code, BB code, and qLDPC codes under both code capacity noise and circuit-level noise. By comparison to the standard decoding algorithms, including MWPM and BPOSD, we have demonstrated the superiority of our approach in accuracy and efficiency. 

Within the generative decoding framework, we have experimented with two types of autoregressive neural networks, MADE and Transformer, and found that MADE with fewer parameters performs better. We suspect that we did not invest enough engineering effort to fully unlock the potential of the Transformer, which is considered the state-of-the-art autoregressive model and has demonstrated its power in large language models. 

Regarding the decoding speed, the time for decoding in our current implementation is in the order of $10^{-2}$ seconds. We attribute this to the relatively low GPU efficiency in our implementation, given that the number of parameters, ranging from $10^6$ to $10^7$, is still comparatively small when compared with large-scale neural network models. As a consequence, the decoding speed is considerably slower than the operation time of the superconducting qubits, which is about $10^{-6}$ seconds per round. So it remains a challenge for GND to improve the efficiency through engineering efforts.
In future work, we aim to enhance the performance and scalability of our generative decoding framework by engineering efforts such as leveraging multiple GPUs, decreasing the latency using FPGAs, employing lower-precision floating-point numbers in decoding, and reducing model size using pruning, distillation, or quantization.

\section{Method}

\subsection{The ELS decomposition and the maximum likelihood decoding}\label{els}
The Pauli group of logical qubits $P_{L}$ is generated by $\mathcal L=\langle l_1^{x},l_1^{z},l_2^{x},l_2^{z},\cdots,l_k^{x},l_k^{z}\rangle$, where $l_i^x$ and $l_i^z$ denote the logical X and Z operators of the $i$'th logical qubits, respectively. 
This group commutes with the stabilizer group. 
Beyond the $2^m$ stabilizer operators and $4^k$ logical operators, there exist another $2^m$ operators that do not commute with the stabilizers; these belong to the pure error subgroup $\mathcal E$, which is Abelian and satisfies the commutation relation $e_ig_j=(-1)^{\delta_{ij}}g_je_i$. 
The three subgroups introduced here elucidate the structure of the Pauli group, as manifested in the decomposition $\mathcal P_n=\mathcal E\otimes \mathcal L\otimes \mathcal S$ \cite{Gottesman1997, gaugephd}, and is termed as $\{\mathcal E,\mathcal L,\mathcal S\}$ decomposition.

The $\{\mathcal E,\mathcal L,\mathcal S\}$ decomposition of the Pauli group tells us that an error $E$ can be mapped to a configuration $\{\alpha,\beta,\gamma\}$ and the error distribution (from the error model) can be equivalently expressed as $p(E)=p(\alpha,\beta,\gamma)$. 
Here $\alpha\in\{0,1\}^m$ represents the configuration of $m$ stabilizers, with each value $\alpha_i$ determined by the commutation relation between $E$ and the pure error generator $E_i$.
The configuration $\beta\in\{0,1\}^{2k}$ denotes the state of the logical $X$ and logical $Z$ operators, while $\gamma\in\{0,1\}^m$ represents the configuration of the $m$ pure error generators, with each value $\gamma_i$ determined by the commutation relation between $E$ and the stabilizers. 
So the configuration $\gamma$ is synonymous with the syndrome.
Utilizing this mapping, we can observe the degeneracy of errors; that is, for a given logical configuration $\beta$, there are $2^m$ assignments of $\{\alpha\}$ that yield the same syndrome. 
MLD computes the likelihood of a logical operator configuration $\beta$, which necessitates considering all $\alpha$ configurations

Consider a $[[n,k,d]]$ quantum code with $n$ physical qubits and distance $d$. A logical state $\ket{\phi}$ with $k$ logical qubits is encoded using a codeword $\ket{\psi}$.

\begin{equation}
\arg\max_{\beta} p(\beta|\gamma) = \arg\max_{\beta}\sum_{\alpha} p(\alpha,\beta|\gamma)\;.
\label{eq:5}
\end{equation}

Thus, in the language of the $\{\mathcal E,\mathcal L,\mathcal S\}$ decomposition, MLD considers the total probability of a coset of the stabilizer sub-group $\mathcal S$, rather than the probability of a single error as in minimum-weight decoding~\cite{Dennis2001, higgott2021pymatching}. This problem generally belongs to the computational class of $\#P$ problems~\cite{Chubb_2021}, meaning that no algorithms can exactly solve MLD in polynomial time. Specifically, the computation poses two challenges: summing over an exponential number of $\alpha$ configurations and performing this summation for the $4^k$ possible $\beta$ configurations~\cite{Bravyi2014}.


\begin{figure*}[!htb]
\centering
{\includegraphics[width=0.46\linewidth]{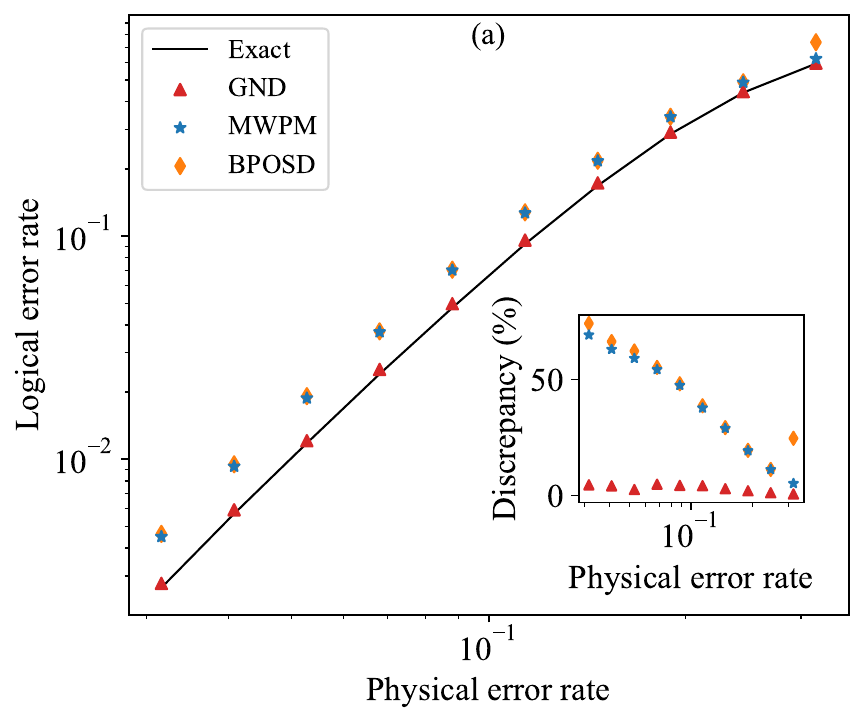}}
{\includegraphics[width=0.46\linewidth]{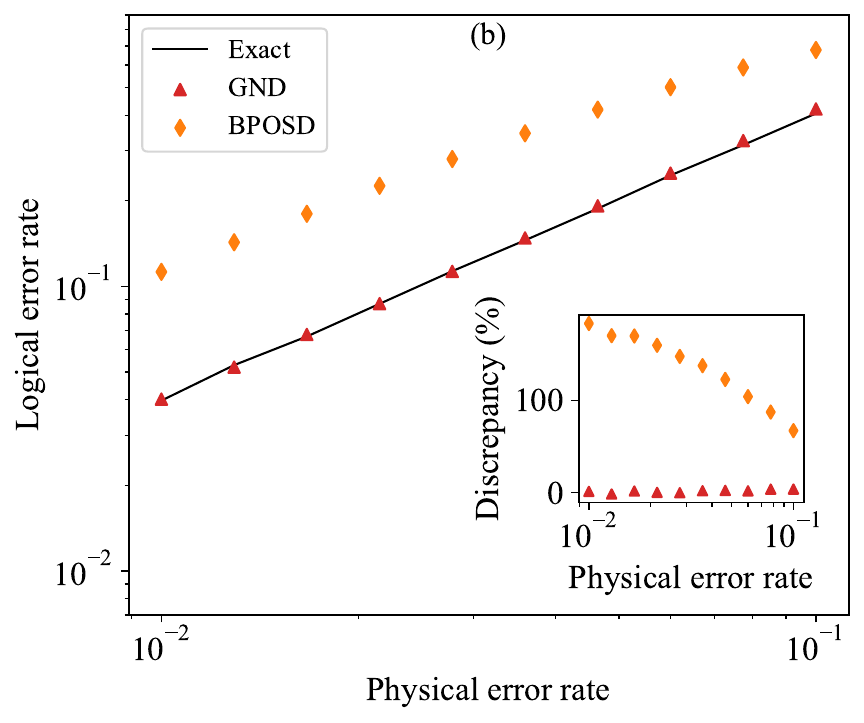}}
\caption{Logical error rates of our algorithm GND with different physical error rates. (a) Compared with MWPM, the exact MLD algorithm, and
BPOSD algorithm on the [[25, 1, 5]] surface code. (b) Compared with an exact MLD algorithm, and
BPOSD algorithm on [[18, 4, 4]] BB code. The error model is the depolarizing error model. Each data point in the figures is averaged over $1000000$ error instances. The black lines are the optimal maximum-likelihood decoding algorithm, which exactly sums all stabilizer configurations. The insets show the discrepancy between the approximate algorithms and the exact algorithm.} 
\label{fig:suface_code}
\end{figure*}

\subsection{Comparing with existing neural network decoder}
Recently, numerous neural network-based decoders (NN decoders) have been proposed alongside the rapid development of artificial intelligence. We can mainly group the NN decoders into two categories, the neural network minimum-weight decoders (NNMWD) and the neural network maximum likelihood decoder (NNMLD).
The NNMWD infers the most likely error for given syndromes by modeling the error probabilities of each data qubits~\cite{8880492, Krastanov2017, PhysRevLett.119.030501, PhysRevLett.122.200501}. One issue for NNMWD is that the output errors are not always consistent with the syndrome~\cite{8880492,PhysRevLett.119.030501, PhysRevLett.122.200501}. This is very similar to the issue of belief propagation which usually does not give an error that satisfies the syndrome in quantum codes, hence requiring an additional operation like OSD to make the syndrome satisfied~\cite{412683, 924874, Roffe2020, Panteleev2021degeneratequantum}.  Compared with the existing NNMWD, our algorithm does not have the syndrome-consistency issue, and the performance of our algorithm (which belongs to MLD) is theoretically better than that of NNMWD.

The second category, NNMLD, learns the joint distribution of logical sectors given syndrome $q(\beta |\gamma)$ using neural networks. For $k=1$, when the number of logical sectors is small, the problem can be converted to a classification problem and solved using a neural network classifier. A successful example is the recently proposed {\textsc{AlphaQubit}}~\cite{bausch_learning_2024} which uses a recurrent neural network and Transformer as a binary classifier (a classifier with two distinct classes for two logical sectors)~\cite{bausch_learning_2024}, which successfully solvers the decoding problem of the surface code up to distance $7$ under circuit level noise~\cite{googlenew2024}.
In this way, the classifier directly models the joint distribution of all possible logical sectors (e.g., using the softmax function) given syndrome. Then we can see that a significant issue of training a classifier is that the number of classes increases with the number of logical sectors $4^k$~\cite{, PhysRevResearch.2.033399} and the computational cost for molding the joint distribution of all $4^k$ logical sectors is not scalable with the number of qubits $k$.

To resolve this issue, several works proposed to simplify the joint distribution of $4^k$ logical sectors using an approximation of the product of marginal distributions~\cite {Varsamopoulos17, Baireuther2017, lange2023datadriven} 
\begin{equation}
      p(\beta_1,\cdots,\beta_{2k}|\gamma) \approx p(\beta_1|\gamma)\cdots p(\beta_{2k}|\gamma)\;.
\end{equation}
In other words, this approach approximates a multi-classification task with $4^k$ classes as $2k$ independent binary-classification tasks.
We term this approach marginal neural decoding (MND). 
In statistical physics, the ansatz of using the product of marginals as an approximation to the joint distribution is known as na\"ive mean-field approximation~\cite{yedidia2003understanding, Wu_2019}, and it has been explicitly shown that the autoregressive distribution (as in our generative approach) can provide achieves much better performance in solving statistical mechanics problems~\cite{Wu_2019}. In machine learning, it is well known that the product ansatz can not provide representative representations for complex distributions such as distributions of words in natural languages~\cite{bishop2023deep}, and modern large language models~\cite{openai2023gpt4} adopt autoregressive distributions (such as in causal Transformers~\cite{Vaswani2017}) rather than product distributions.

\begin{figure}
    \centering
    \includegraphics[width=0.9\linewidth]{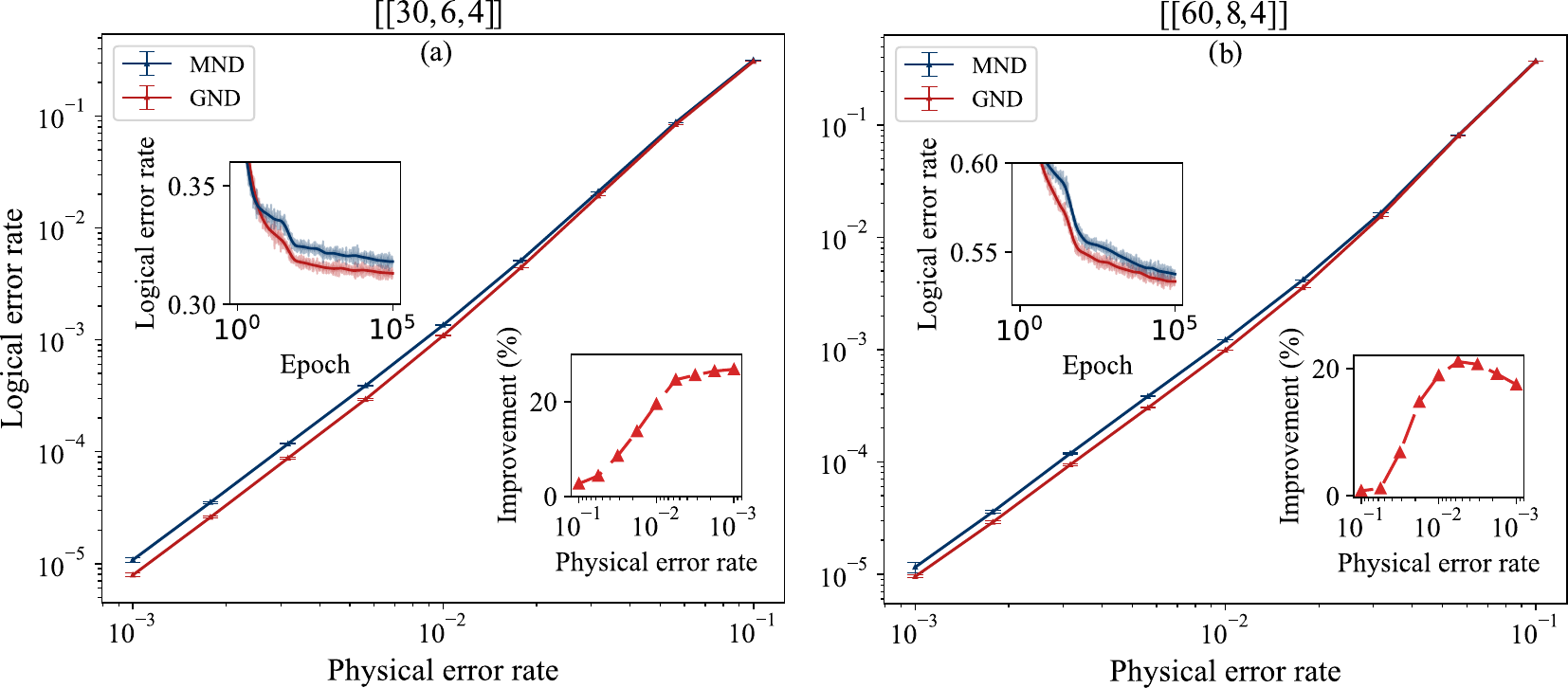}
    \caption{Comparing GND and MND. The sub-figures (a) and (b) show the decoding performance of GND and MND on [[30, 6, 4]] and [[60, 8, 4]] codes, respectively. The lower right insert figures show the improvement of GND relative to MND. And the upper left insert figures show the variation of the logical error rate with training epoch. The red and blue solid lines represent the average logical error rate of GND and MND, respectively, and shades of the same color indicate variance. MADE and MLP are implemented to construct the decoders. The number of parameters used in GND is 1980972 and 11666828 for [[30, 6, 4]] and [[60, 8, 4]] codes separately. These numbers are 1993600 and 11557728 of MND. The logical error rate with epoch of the [[30, 6, 4]] code is shown in the inset. The decoders are trained at $p=0.1$ for [[30, 6, 4]] code and $p=0.12$ for [[60, 8, 4]]. Each data point averages over 10 decoding tasks and each task has $10^7$ shots.} 
    \label{fig:2}
\end{figure}

In the context of the neural network decoder, we explicitly compared the performance of our generative neural decoding GND approach with the neural network decoder using the product of marginals MND. In detail, we performed numerical experiments on [[30, 6, 4]] and [[60, 8, 4]] high-rate codes and compared the logical error rates given by GND and MND. In both GND and MND, we explored the same neural network architecture, using MADE~\cite{germain2015made} and Multi-Layer Perceptron (MLP)~\cite{Varsamopoulos17} so the parameters of them are quite similar. The results are shown in Fig.~\ref{fig:2}. From the figure, we can see that GND demonstrates significant advantages over MND, indicating that the principle, the autoregressive modeling, is significantly superior to the simple product of marginals as used in previous MND studies.

We observe that another notable distinction between our generative decoder and existing neural network decoders is that our generative decoder models the joint probability \(p(\beta, \gamma)\), as illustrated in Eq.~\eqref{eq:3}. This implies that the generative decoder not only models \(p(\beta \mid \gamma)\) but also \(p(\gamma)\). Consequently, in addition to generating logical sectors, it also has the capability to generate syndromes.

\subsection{Comparing MADE with Transformer in decoding}
\label{sec:Iv.c}
\begin{figure}[htbp!]
    \centering
    \includegraphics[width=0.5\linewidth]{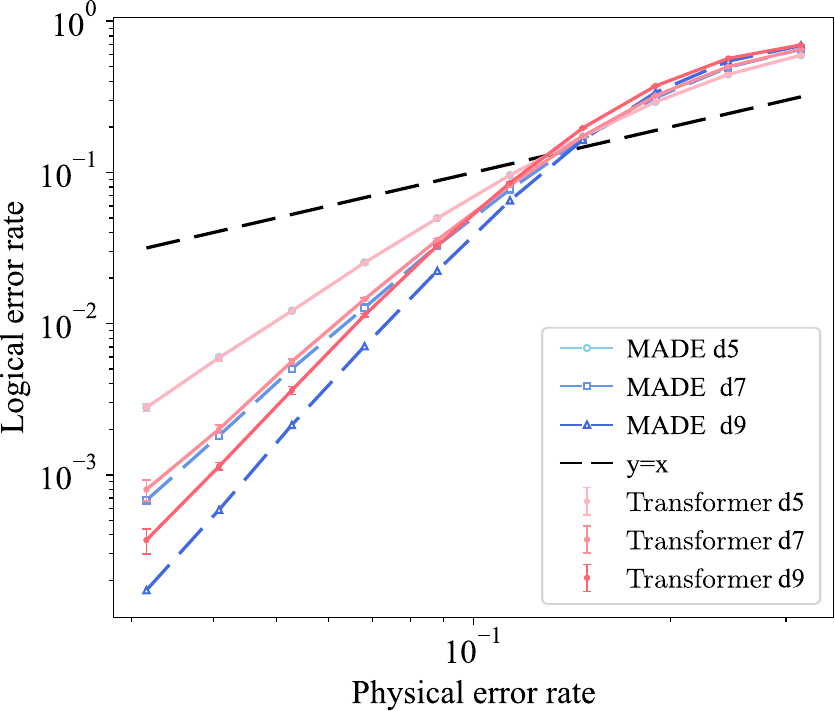}
    \caption{Comparing the decoding performance of MADE with Transformer on $d=[3, 5, 7]$ Rotated Surface Codes. In this figure, we use the hyperparameters of Transformer as $\mathrm{d}_{model}=256$, $\mathrm{N}_{head}=4$, $\mathrm{d}_{ff} =256$, $\mathrm{N}_{layers}=2$. And the hyperparameters of MADE can be found in Tab.~\ref{tab:1}. Each data point has 10 decoding tasks, and each task has $10^5$ shots.}
    \label{fig:5}
\end{figure}

In this section, we evaluated two different autoregressive models in our GND framework, MADE~\cite{germain2015made} and causal Transformer~\cite{fakoor2020trade}. The architectures of these models are illustrated in Fig.~\ref{fig:1}. MADE is a lightweight autoregressive model composed exclusively of masked linear layers and activation functions, enforcing the autoregressive property through binary connectivity masks. Its simplicity and efficiency make it a popular choice for probabilistic modeling in statistical physics, particularly for partition function estimation~\cite{PhysRevLett.122.080602}. In contrast, the causal transformer leverages the decoder module of the Transformer architecture~\cite{Vaswani2017}, which has achieved state-of-the-art performance in language modeling tasks~\cite{chatGPT}. Despite its success in other domains, the transformer underperformed MADE in our quantum decoding framework. This may be due to the large number of parameters of the transformer that are difficult to optimize, as well as the self-attention mechanism, which dynamically models dependencies between variables, may struggle with problems where variable relationships are unknown or poorly structured, as noted in previous work~\cite{10.1162/tacl_a_00306}. 
To quantify this disparity, Fig.~\ref{fig:5} compares the decoding accuracy of MADE and Transformer on increasing size rotated surface codes. While both models perform similarly for small code distances, the Transformer's performance degrades significantly as the system scales. In addition, its memory footprint exceeds the capacity of a single NVIDIA A100 GPU at larger scales, when the code distance is up to 11. We conducted extensive studies to isolate the cause of the Transformer's shortcomings and tried to remove modules that might affect performance, such as dropout and layer norm, among others. All these attempts failed to improve the performance of the Transformer any further. 

\subsection{Parameters of the neural networks}
In the numerical experiments, all neural networks were trained on a single NVIDIA A100 GPU. The hyperparameters for training the neural networks are detailed in Tab.~\ref{tab:1}. The table provides the number of inputs $N_{in}$, Depth, Width and number of parameters $N_{\text{para}}$ of neural networks, across multiple error-correcting codes, including RSC (Rotated Rurface codes), LDPC (low-density parity-check codes), BB (Bivariate Bicycle codes) and $\mathrm{DSC}^{\mathrm{cir}}$ (Defected Surface codes with circuit level noise). $P_t$ is the physical error rate, $T_{\mathrm{decoding}}$ is the decoding time in seconds.

\begin{table}[!htbp]
\setlength{\tabcolsep}{3pt}
    \centering
    \begin{ruledtabular}
    \begin{tabular}{ccccccc}
    \textrm{Code}&
    \textrm{$N_{in}$}&
    \textrm{Depth}&
    \textrm{Width}&
    \textrm{$P_t$}&
    \textrm{$N_{\text{para}}$}&
    \textrm{$T_{\mathrm{decoding}}$}\\
    \hline
    RSC7 & $50$ & $4$ & $30$ & 0.189 & 3522050&$8.5\times10^{-4}$\\
    RSC9 & $82$ & $4$ & $30$ & 0.189 & 9397282&$8.5\times10^{-4}$\\
    RSC11 & $122$ & $4$ & $20$ & 0.189 & 9308722&$1.1\times10^{-3}$\\
    RSC13 & $170$ & $4$ & $25$ & 0.189 & 27988545&$9.2\times10^{-4}$\\
    LDPC30 & $36$ & $4$ & $20$ & 0.1 & 1980972&$4\times10^{-3}$\\
    BB60 & $68$ & $5$ & $30$ & 0.12 & 11666828 &$6.4\times10^{-3}$\\
    BB72 & $84$ & $6$ & $30$ & 0.12 & 16289364 &$9.3\times10^{-3}$\\
    $\mathrm{DSC}^{\mathrm{cir}}_{k2}$&191&4&30&0.015&66789071&$1.2\times10^{-3}$\\
    $\mathrm{DSC}^{\mathrm{cir}}_{k3}$&189&4&30&0.015&65397969&$1.5\times10^{-3}$\\
    $\mathrm{DSC}^{\mathrm{cir}}_{k4}$&187&4&30&0.015&64021507&$1.9\times10^{-3}$\\

    \end{tabular}
    \end{ruledtabular}
    \caption{Hyperparameters of GND. Where $N_{in}$ is the number of inputs of the model, $Depth$ is the number of hidden layers, $width$ is the number of channels, $P_t$ is the physical error rate for training, $N_{para}$ is the number of parameters and $T_{\mathrm{decoding}}$ is the decoding time in seconds. }
    \label{tab:1}
\end{table}

\begin{acknowledgments}
A \textsc{python} implementation and a Jupyter Notebook tutorial of our algorithm are available at \cite{github}.
    We thank Weilei Zeng and Ying Li for their helpful discussions. We are also grateful to the anonymous reviewers for their valuable suggestions during the early stages of this paper.
    This work is supported by Project  12325501, 12047503, and 12247104 of the National Natural Science Foundation of China and Project ZDRW-XX-2022-3-02 of the Chinese Academy of Sciences. P.~Z. is partially supported by the
Innovation Program for Quantum Science and Technology
project 2021ZD0301900.
\end{acknowledgments}

\section*{Author contributions}
H.C., F.P., and P.Z. had the original idea for this work. H.C., F.P., D.F., and Y. W. performed the study, and all authors contributed to the preparation of the manuscript.

\section*{Competing interests}

The authors declare no competing interests.

\section*{Additional information}



\textbf{Correspondence and requests for materials} should be addressed to P.Z.

\textbf{Reprints and permission information}  is available online at  [URL will be inserted by publisher].

\bibliography{gnd}

\newpage
\appendix


\end{document}